\documentclass[a4paper,11pt]{article}

\usepackage{verbatim}
\usepackage{xspace}
\usepackage{multicol}

\newtheorem{defin}{Definition}[section]

\newcommand{\system}{\S=\langle\K,\R,\M\rangle}
\newcommand{\MC}{M_c}
\newcommand{\wrt}{w.r.t.\ }

\newcommand{\ans}{\mathit{ans}}

\newcommand{\nt}[1]{\langle\mathit{#1}\rangle}

\newcommand{\inclusion}[2]{#1\sqsubseteq #2}
\newcommand{\disjoint}[2]{#1\ \texttt{disj}\ #2}
\newcommand{\functional}[2]{(\texttt{funct}\ \exists[#1]#2)}

\newcommand{\inclass}{\inclusion{B_1}{B_2}}
\newcommand{\disjass}{\disjoint{B_1}{B_2}}
\newcommand{\functass}{\functional{i}{R}}

\newcommand{\A}{\mathcal{A}}
\newcommand{\I}{\mathcal{I}}
\renewcommand{\S}{\mathcal{S}}
\newcommand{\K}{\mathcal{K}}
\newcommand{\D}{\mathcal{D}}
\newcommand{\M}{\mathcal{M}}
\newcommand{\R}{\mathcal{R}}

\newcommand{\dllite}{\textit{DL-Lite}\xspace}
\newcommand{\dlrdb}{\textit{DLR-DB}\xspace}
\newcommand{\dlrlite}{\textit{DLR-Lite}\xspace}

\title{Efficient Query Answering over Conceptual Schemas of Relational
  Databases\thanks{This technical report in anonymous form was accompanying a
    submission to the student session of ESSLII'06. It was written by Lina
    Lubyt{}\.{e}, \v{Z}ivil{}\.{e} Nork\={u}nait{}\.{e}, Mantas \v{S}imkus and
    Daniel Trivellato as part of the Master student project carried out at the
    Free University of Bolzano (Italy) under the supervision of Diego
    Calvanese and Sergio Tessaris. Evaldas Taroza has contributed in preparing
    the submission to ESSLII'06.}} \date{}

\begin{document}
\maketitle

\begin{abstract}
We develope a query answering system, where at the core of the work there is an idea of query answering by rewriting. For this purpose we extend the DL \dllite [5] with the ability to support n-ary relations, obtaining the DL \dlrlite, which is still polynomial in the size of the data [3,4]. We devise a flexible way of mapping the conceptual level to the relational level, which provides the users an SQL-like query language over the conceptual schema. The rewriting technique adds value to conventional query answering techniques, allowing to formulate more simple queries, with the ability to infer additional information that was not stated explicitly in the user query. The formalization of the conceptual schema and the developed reasoning technique allow checking for consistency between the database and the conceptual schema, thus improving the trustiness of the information system.

\end{abstract}

\section{Introduction}
The research we are currently carrying out is aimed at the development of a
query answering system that enables users to pose queries over the
conceptual schema of a database. Such a system provides added value against
conventional DBMSs, where the users are exposed the relational schema only. At
the core of our work there is an idea of query answering by rewriting.

In general, query answering by rewriting is divided into two phases. The
first one re-expresses a user query posed over the conceptual schema in
terms of the relations at the underlying database, and the second evaluates
the rewriting over the underlying database (e.g.,[1]).

Our approach uses a formalism based on Description Logics (DLs) [2]
to formalize the conceptual schema of the database. Specifically, we have
extended the DL \dllite [5] with the ability to support n-ary
relations, obtaining the DL \dlrlite.  Such a formalism is expressive enough
to capture basic Entity-Relationship or UML Class diagrams, while allowing
query answering that fully takes into account the constraints in the
conceptual schema and is still tractable (i.e., polynomial) in the size of
the data [3,4].

We have devised a flexible way of mapping the conceptual level to the
underlying relational level, which provides the users an SQL-like query
language over the conceptual schema. Queries at the conceptual level are
first translated into the relational level queries by taking into account
the mapping of entities and relationships to the actual database relations.
To provide a complete answer to the query, the system then uses the
developed query rewriting technique to take into account the constraints
expressed in the conceptual schema. The initial user query is thus
translated to a set of SQL queries that are evaluated by the DBMS.

This rewriting technique adds value to conventional query answering
techniques.  Firstly, the user is allowed to formulate more simple queries
using terms defined in the conceptual schema only, without taking into
account some relational database related details (e.g., join attributes).
Moreover, the query rewriting technique allows one to infer additional
information that was not stated explicitly in the user query but is implied
by the constraints at the conceptual level. Last but not least, the
formalization of the conceptual schema and the developed reasoning technique
allow checking the consistency of the underlying database against the
conceptual schema, therefore, the trustiness of the information system is
improved.
\section{Formal Framework}

\dlrdb system is a triple $\system$, where $\K$ is the knowledge base (KB) of $\S$, $\R$ is a relational schema for $\S$ and $\M$ is the mapping between the KB $\K$ and the relational schema $\R$. 

\subsection{Conceptual Level}

We call our description logic language \dlrlite, that allows to represent the domain of interest in terms of concepts, denoting sets of objects, and relationships, denoting relations between objects. In the language, \emph{basic concepts} are defined as follows: 
\begin{displaymath}
B \ ::=\ A\ |\ \exists{[i]R}																																														\end{displaymath}
where $A$ denotes an atomic concept, $R$ an $n$-ary relationship, and $1\leq i \leq n$. Intuitively, $\exists{[i]R}$ denotes the projection of $R$ on the \emph{i-th} component. Note, that all concepts denote unary predicates.

For representing intensional knowledge in the KB, we have assertions of the form:
\begin{displaymath}  
\inclass \qquad \qquad \emph{(inclusion)} 
\end{displaymath}
\begin{displaymath} 
\disjass  \qquad \qquad \emph{(disjointness)}
\end{displaymath}
\begin{displaymath}  
\functass \qquad \qquad \emph{(functionality)}
\end{displaymath}

An inclusion assertion expresses that a basic concept is subsumed by another concept, a disjointness assertion states that the set of objects denoted by a basic concept $B_1$ is disjoint from the ones denoted by another concept $B_2$, while a functionality assertion expresses the (global) functionality of a certain component of a relationship.

The formal meaning of concept descriptions above is given in terms of interpretations over a fixed infinite countable \emph{domain} $\Delta$. We assume, we have one constant for each object, denoting exactly that object.

An \emph{interpretation} $\I = (\Delta, \cdot^{\I})$ consists of a first order structure over $\Delta$ with an \emph{interpretation function} $\cdot^{\I}$ such that: 
\begin{displaymath}
A^{\I} \subseteq \Delta 
\end{displaymath}
\begin{displaymath}
R^{\I} \subseteq \Delta ^{n}
\end{displaymath}
\begin{displaymath}
(\exists{[i]R})^{\I} = \{ \ c\ | \ \exists (c_1,\ldots , c_n) \in R^{\I}, \ c = c_i \}.
\end{displaymath}

An interpretation $\I$ \emph{satisfies} an inclusion assertion $\inclass$ iff $B_1^{\I} \subseteq B_2^{\I}$; $\I$ satisfies a disjointness assertion $\disjass$ iff $B_1^{\I} \ \cap B_2^{\I} \ = \ \emptyset$; $\I$ satisfies a functionality assertion $\functass$ if $(c_1, \ldots ,c_i, \ldots c_n) \in R^{\I} \land (c'_1,\ldots, c_i, \ldots ,c'_n) \in R^{\I} \supset c_1=c'_1,\ldots ,c_n=c'_n$.

A \emph{model of a KB} $\K$ is an interpretation $\I$ that satisfies all the assertions in $\K$. A KB is \emph{satisfiable}, if it has at least one model. A KB $\K$ \emph{logically implies} an assertion $\alpha$ if all the models of $\K$ satisfy $\alpha$.

All presented assertions allow us to specify the typical constructs used in conceptual modeling. Specifically:
\begin{itemize}
\item[-] \emph{ISA}, using assertions of the form $\inclass$, stating that the
  class $B_1$ is a subclass of the class $B_2$;
\item[-] \emph{class disjointness}, using assertions of the form $\disjass$, stating disjointness between the two classes $B_1$ and $B_2$;
\item[-] \emph{role-typing}, using assertions of the form $\exists{[i]R}\sqsubseteq B$, stating that the \emph{i-th}
  component of the relationship $R$ is of type $B$;
\item[-] \emph{participation constraints}, using assertions of the form $B\sqsubseteq\exists{[i]R}$, stating that instances of class $B$ participate to the relationship $R$ as the \emph{i-th} component;
\item[-] \emph{non-participation constraints}, using assertions of the form $B \ \texttt{disj} \ \exists{[i]R}$, stating that instances of class $B$ do not participate to the relationship $R$ as the \emph{i-th} component;
\item[-] \emph{functionality restrictions}, using assertions of the form $\functass$, stating that an object can be the \emph{i-th} component of the relationship $R$ at most once.
\end{itemize}

\paragraph{Example 1} Consider atomic concepts \emph{Student}, \emph{Professor} and \emph{Course}, the relationships \emph{Attends} between \emph{Student} and \emph{Course}, \emph{Teaches} between \emph{Professor} and \emph{Course}, and \emph{HasTutor} between \emph{Student} and \emph{Professor}. We can now define the following inclusion, disjointness and functionality assertions:
\begin{multicols}{2}
\begin{enumerate}
\item[($A_1$)] $\exists{[1]Attends}\sqsubseteq Student$
\item[($A_2$)] $\exists{[2]Attends}\sqsubseteq Course$
\item[($A_3$)] $\exists{[1]Teaches}\sqsubseteq Course$
\item[($A_4$)] $\exists{[2]Teaches}\sqsubseteq Professor$
\item[($A_5$)] $Professor \sqsubseteq \exists{[2]Teaches}$
\item[($A_6$)] $\exists{[1]HasTutor}\sqsubseteq Student$
\item[($A_7$)] $\exists{[2]HasTutor}\sqsubseteq Professor$
\item[($A_8$)] $Student \sqsubseteq \exists{[1]Attends}$
\item[($A_9$)] $Student \sqsubseteq \exists{[1]HasTutor}$
\item[($A_{10}$)] $Course \sqsubseteq \exists{[2]Attends}$
\item[($A_{11}$)] $Course \sqsubseteq \exists{[1]Teaches}$
\item[($A_{12}$)] $(\texttt{funct} \ \exists{[1]HasTutor})$
\item[($A_{13}$)] $(\texttt{funct} \ \exists{[1]Teaches})$
\end{enumerate}
\end{multicols} 
where $A_1$ states that everyone attending a course must be a student, while $A_2$ states that all attended courses has to be only those that are offered in general, etc. $A_{12}$ states that a student can have only one tutor, and $A_{13}$ states that a course can be tought by only one professor.

We denote by \texttt{Normalize}$(\K)$ the \dlrlite KB obtained by transforming
the KB $\K$ as follows. The KB $\K$ is expanded by computing all disjoint
inclusions between basic concepts implied by $\K$. More precisely, the $\K$ is
closed with respect to the following inference rule: if $B_1\sqsubseteq B_2$
occurs in $\K$ and either $\disjoint{B_2}{B_3}$ or $\disjoint{B_3}{B_2}$ occurs
in $\K$, then add $\disjoint{B_1}{B_3}$ to $\K$.

It is immediate to see that, for every \dlrlite KB $\K$, \texttt{Normalize}$(\K)$ is equivalent to $\K$, in the sense that the set of models of $\K$ coincides with that of \texttt{Normalize}$(\K)$.

Given a \dlrdb system $\system$, \texttt{Normalize}$(\S) = \langle \K_n, \R, \M \rangle$, where $\K_n$ = \texttt{Normalize}$(\K)$.

\subsection{Relational Level}

At the relational level we consider \emph{relations}, where each relation has
an associated sequence of typed attributes. Each relation may have a sequence
of one or more \emph{components}, where each component is a sequence of
attributes of the relation. Components may not overlap. We call attributes that
do not belong to any component, \emph{additional attributes} of the relation.
Note, that the order of components and the order of attributes may not
necessarily be related to each other.

\subsection{Mapping from Conceptual to Relational Level}

We can now define the mapping $\M$ between conceptual and logical level as
follows:
\begin{itemize}
\item to each atomic concept $A$, $\M$ associates a relation $\M(A)$ with a
  single component;
\item to each $n$-ary relationship $R$, $\M$ associates a relation $\M(R)$ with
  $n$ components.
\end{itemize}
The mapping induces a \emph{signature} on basic concepts, and specifically
\begin{itemize}
\item for an atomic concept $A$, the signature is the sequence of types of
  attributes of the component of the relation corresponding to $A$.
\item for a concept of the form $\exists{[i]R}$, the signature is the sequence
  of types of the $i$-th component of the relation corresponding to $R$.
\end{itemize}
A mapping $\M$ is \emph{consistent} with the conceptual level $\K$ and the
relational level $\R$ of a system $\system$, if for each inclusion assertion
$\inclass$ in $\K$, the signature of $B_1$ is equal to the signature of $B_2$.
Note that for disjointness assertions $\disjass$, we do not require $B_1$ and
$B_2$ to have the same signature.
Indeed, if $B_1$ and $B_2$ have different signatures, the disjointness
assertions will trivially be satisfied at the relational level.
In the following, we will always assume that in a system $\system$, the mapping
$\M$ is consistent with $\K$ and $\R$.

\paragraph{Example 1 (contd.)} In the table below for all atomic concepts and relationships the mapping associates the corresponding relations with components (underlined) and additional attributes.\\ \\
\begin{tabular}{p{2.5cm}|l}
\hline \hline
\textbf{Concept/ Relationship} & \textbf{Relation}\\
\hline \hline 
Student & StudentTable(\underline{SName, SSurname}, EnrollNumber)\\
Course & CourseTable(\underline{CourseId}, Name, Category)\\
Professor & ProfessorTable (\underline{PName, PSurname}, Degree)\\
Attends & AttendsTable (\underline{SName, SSurname}, \underline{CourseId}, Year)\\
Teaches & TeachesTable  (\underline{PName, PSurname}, \underline{CourseId}, Semester)\\
HasTutor & HasTutorTable (\underline{SName, SSurname}, \underline{PName, PSurname})\\ 
\end{tabular}

\subsection{Semantics of a System $\S$}


In order to define the semantics of a system $\system$, we first extend the
mapping $\M$ to a mapping $\MC$ from basic concepts to components of relations
as follows:
\begin{itemize}
\item for an atomic concept $A$, let $\A$ be the sequence of attributes
  corresponding to the only component of $\M(A)$. Then $\MC(A)=\pi_\A(\M(A))$;
\item for a relationship $R$, let $\A$ be the sequence of attributes
  corresponding to the $i$-th component of $\M(R)$. Then
  $\MC(\exists[i]R)=\pi_\A(\M(R))$.
\end{itemize}

A \emph{database instance} (or simply \emph{database}) $\D$ over the relational
schema $\R$ is the set of facts of the form $R(\vec{c})$, where $R$ is a
relation of arity $n$ in $\R$ and $\vec{c}$ is an $n$-tuple of constants of
$\Delta$.
A database $\D$ satisfies \wrt $\S$
\begin{itemize}
\item an inclusion assertion $\inclass$, if $(\MC(B_1))^\D \subseteq
  (\MC(B_2))^\D$;
\item a disjointness assertion $\disjass$, if $(\MC(B_1))^\D \cap (\MC(B_2))^\D
  = \emptyset$
\item a functionality assertion $\functass$, if the cardinality of
  $(\MC(\exists[i]R))^\D$ is equal to the cardinality of $(\M(R))^\D$. In other
  words, the set of attributes of the $i$-th component of $R$ is a key of
  $R^{\D}$.
\end{itemize}

A database $\D$ is said to be \emph{consistent} \wrt a system $\system$, if it
satisfies \wrt $\S$ all assertions in $\K$. A database $\D$ is said to be
\emph{df-consistent} \wrt $\S$, if it satisfies \wrt $\S$ all disjointness and
functionality assertions in $\K$.


\section{Queries}

\subsection{Queries over Conceptual Level}

Queries over a \dlrdb system $\system$ are specified using an SQL-like syntax
corresponding to SPJ queries. More precisely, such a query is written in the
form:
\begin{quote}
  \texttt{SELECT} $\nt{attribute\_specifications}$\\
  \texttt{FROM} $\nt{relationship\_specifications}$\\
  \texttt{WHERE} $\nt{selection\_conditions}$
\end{quote}
where \begin{itemize} 
\item $\nt{relationship\_specifications}$ denotes the concepts and relationships
  involved in the query and the way they join together. It is defined as
  follows:
  \[
    \begin{array}{l}
      \nt{relationship\_specifications} ::=  \nt{rel\_spec} \mid
        \nt{relationship\_specifications},\nt{rel\_spec}\\
      \nt{rel\_spec}  ::=  \nt{join} \texttt{ ON } \nt{conditions}\\
      \nt{join} ::= \nt{relationship} \mid
        \nt{join} \texttt{ JOIN } \nt{relationship}\\
      \nt{relationship} ::= C_i \texttt{ AS } V_{i}\\
      \nt{conditions} ::=  \nt{equality} \mid
        \nt{conditions} \texttt{ AND } \nt{equality}\\
      \nt{equality} ::= e_i = e_j
    \end{array}
  \]
  
  Intuitively, $\nt{relationship\_specifications}$ is a sequence of expressions
  of one of the following forms:
  \begin{itemize}
  \item $C \texttt{ AS } V$
  \item
    $\begin{array}[t]{@{}l}
      C_1 \texttt{ AS } V_1 \texttt{ JOIN }
      C_2 \texttt{ AS } V_2 \texttt{ JOIN } \cdots \texttt{ JOIN }
      C_k \texttt{ AS } V_k\\
      \quad
      \texttt{ ON } e_1=e_2  \texttt{ AND } \cdots \texttt{ AND } e_{h-1}=e_h
    \end{array}$
  \end{itemize}
  where
  \begin{itemize}
  \item each $C_j$ denotes the name of a relationship or an atomic concept in
    $\K$;
  \item each $V_j$ is a unique variable name, associated to
    $C_j$\footnote{Note that relationships and atomic concepts may be
     repeated.};
  \item in the equalities $e_i=e_j$, each $e_i$ or $e_j$ is either
    \begin{itemize}
    \item $V$, if $V$ is a variable corresponding to an atomic concept;
    \item $V.i$, if $V$ is a variable corresponding to a relationship of arity
      $n\geq i$,
    \end{itemize}
  \item the signatures of the two associated concepts/relationships components
    must be the same.
  \end{itemize}
  
\item $\nt{attribute\_specifications}$ is a sequence of attributes of the form
  $V.a$, where $V$ is a variable in $\nt{relationship\_specifications}$,
  associated to concept or relationship $C$, and $a$ is an attribute of
  relation $\M(C)$;
\item $\nt{selection\_conditions}$ is a set of equalities, each of one of the
  following forms:
  \begin{itemize}
  \item $V_1.a_1=V_2.a_2$,
  \item $V_1.a_1=c$,
  \end{itemize}
  where $V_i$ is a variables in $\nt{relationship\_specifications}$, associated
  to concept or relationships $C$, $a_i$ is an attribute of relation $\M(C)$,
  and $c$ is a constant.
\end{itemize}

\subsection{Conjunctive Queries over the Relational Level}

In this section we first recall the notion of a conjunctive query (CQ).
Afterwards we present how a CQ over the relational level can be obtained from a
query over the conceptual level.

\subsubsection{Conjunctive Queries}

A \emph{term} is either a variable or a constant. An \emph{atom} is an
expression $p(z_1, \ldots, z_n)$, where $p$ is a predicate (relation) of arity
$n$ and $z_1, \ldots, z_n$ are terms.  A conjunctive query \emph{q} over a
knowledge base $\K$ is an expression of the form
\[
  q(\vec{x}) \ \leftarrow \ \exists{\vec{y}}. conj(\vec{x}, \vec{y})
\]
where $\vec{x}$ are the so-called \emph{distinguished variables}, $\vec{y}$ are
existentially quantified variables called \emph{non-distinguished variables},
and $conj(\vec{x}, \vec{y})$ is a conjunction of atoms of the form $T(z_1,
\ldots , z_n)$, where $T$ is a relation of $\R$ with $n$ attributes and $z_1,
\ldots , z_n$ are terms. $q(\vec{x})$ is called the \emph{head} of $q$ and
$\exists{\vec{y}}. conj(\vec{x}, \vec{y})$ the \emph{body} of $q$.

The \emph{answer} of a query $q(\vec{x}) \ \leftarrow \
\exists{\vec{y}}.conj(\vec{x}, \vec{y})$ over a database $\D$ is the set
$q^{\D}$ of tuples $\vec{c}$ of constants in a domain $\Delta$ such that when
we substitute the variables $\vec{x}$ with the constants $\vec{c}$, the formula
$\exists{\vec{y}}.conj(\vec{x}, \vec{y})$ evaluates to true in $\D$.

A \emph{union of conjunctive queries (UCQ)} is an expression
\[
q(\vec{x}) \ \leftarrow \ \exists{\vec{y_1}}. conj_1(\vec{x}, \vec{y_1}) \lor
\cdots \lor \exists{\vec{y_m}}. conj_m(\vec{x}, \vec{y_m})
\]
where for each $i \in \{ 1, \ldots , m\} \ conj_i(\vec{x}, \vec{y_i})$ is a
conjunction of atoms.

The answer of a UCQ $q(\vec{x}) \ \leftarrow \
\exists{\vec{y_1}}.conj_1(\vec{x}, \vec{y_1}) \lor \cdots \lor
\exists{\vec{y_m}}.conj_m(\vec{x}, \vec{y_m})$ over a database $\D$ is the
union of the answers of the conjunctive queries
\[
  \begin{array}{rcl}
    q_1(\vec{x}) &\leftarrow& \exists{\vec{y_1}}. conj_1(\vec{x}, \vec{y_1})\\
    &\vdots&\\
    q_m(\vec{x}) &\leftarrow& \exists{\vec{y_m}}. conj_m(\vec{x}, \vec{y_m})
  \end{array}
\]

\subsubsection{Converting Conceptual Queries to Conjunctive Queries} 

Given a \dlrdb system $\system$, the conversion of a query $q$ over the
conceptual level into a conjunctive query is done in two steps:
\begin{enumerate}
\item the query $q$ is converted into a standard SQL select-project-join
  query $q'$ over the relational schema $\R$;
\item $q'$ is converted into a conjunctive queries using the standard
  translation.
\end{enumerate}
In this conversion, the order of attributes of a relation $R$, specified at the relational level, is preserved in the atoms for $R$.

In order to convert our conceptual queries to standard SQL queries, first each
relationship $C_j$ is substituted with $\M(C_j)$. For each equality $e_1=e_2$
in the conceptual query, we substitute it with the conjunction of equalities
between the attributes corresponding to the components mentioned in $e_1$ and
$e_2$.

\paragraph{Example 1 (contd.)} Suppose we want to know the surnames of all students that attend the course with ID ''AB23INF''. We formulate the conceptual query as follows: 
\begin{verbatim}
  SELECT S.Surname
  FROM Student AS S JOIN Attends AS A ON S = A.1
  WHERE A.Course = "AB23INF"
\end{verbatim}
After the rewriting we get the following SQL query:
\begin{verbatim}
  SELECT S.Surname
  FROM StudentTable AS S JOIN AttendsTable AS A
       ON S.Name = A.Name AND S.Surname = A.Surname
  WHERE A.Course = "AB23INF"
\end{verbatim}

Given a query $q$ over $\S$, we denote with $CQ(q, \S)$ the conjunctive query over
$\R$ resulting from the above conversion.

In order to evaluate, using a relational DBMS, the queries we get from the rewriting procedure, we need to convert them back to SQL. In doing so, we again make use of the order of attributes specified at the relational level. We denote the conversion of a CQ $q$ to SQL with $SQL(q, \R)$.

\subsection{Reasoning in \dlrdb system $\S$}

Given a \dlrdb system $\system$, a conceptual query $q$ over $\S$ and a
database $\D$ over $\R$, the \emph{certain answers} $\ans(q,\S,\D)$ is the set
of tuples $\vec{c}$ of constants of $\Delta$, such that $\vec{c} \in
q_{\S}^{\D'}$ for every database $\D'$ that includes $\D$ and is consistent
with $\S$.
 
The basic reasoning services over a \dlrdb system $\system$ are:
\begin{itemize}
\item \emph{KB satisfiability}: verify whether a KB is satisfiable.
\item \emph{query answering}: given a \dlrdb system $\system$, a conceptual
  query $q$ over $\S$ and a database $\D$ over $\R$, return the certain answers
  ans$(q, \S, \D)$.
\item \emph{query rewriting}: given a \dlrdb system $\system$, and a conceptual
  query $q$ over $\S$, return a query $q_r$ over $\R$, such that
  $q_r^{\D}=ans(q, \S, \D)$ for every database $\D$ that is df-consistent with
  \texttt{Normalize}$(\S)$.
\end{itemize}


\section{Query Rewriting in System $\S$}

In this section we present an algorithm that computes the perfect rewriting of a UCQ. Before proceeding, we address some preliminary issues. 

\paragraph{df-consistency of $\D$ w.r.t. $\S$} The algorithm \texttt{Consistent} takes as input a normalized KB $\K$ and verifies the following conditions:
\begin{itemize}
\item[-] if there exists a disjunction assertion $\disjass$, such that $(\M_C(B_1))^\D \cap (\M_C(B_2))^\D \neq \emptyset$ 
\item[-] if there exists a functionality assertion $\functass$, such that the cardinality of $(\M_C(\exists[i]R))^\D$ is not equal to the cardinality of $(\M(R))^\D$.
\end{itemize}

Informally, the first condition corresponds to checking whether $\D$ explicitly contradicts some disjunction assertion in $\K$, and the second condition corresponds to check whether $\D$ violates some functionality assertion in $\K$. If at least one of the above conditions holds, then the algorithm returns \emph{false}, i.e., $\D$ is not fk-consistent w.r.t. $\S$. Otherwise, the algorithm returns \emph{true}.

\subsection{Rewriting}

The basic idea of the method used is to reformulate the query taking into account the KB $\K$ [4]: in particular, given a query $q$ over the conceptual schema $\K$, we compile the assertions of the KB into the query itself, thus obtaining a new query $q'$. Such a new query is then evaluated over the database instance $\D$.

We say that an argument of an atom in a query is \emph{bound} if it corresponds to either a distinguished variable or a shared variable, i.e., a variable occurring at least twice in the query body, or a constant, while we say that it is \emph{unbound} if it corresponds to a non-distinguished non-shared variable.

\begin{defin} We indicate with gr(g, I) the atom obtained from the atom g by applying the inclusion assertion I as follows:\\
an inclusion assertion $B \sqsubseteq A$ (resp. $B \sqsubseteq \exists[i]R$) is applicable to an atom $T(x_1, \ldots, x_n)$ if
\begin{itemize}
\item[(i)] $\M(A)=T$ (resp. $\M(R)=T$)
\item[(ii)] all variables among $x_1, \ldots, x_n$ that are in positions of $T$ that are not part of the only (resp. the $i-th$) component of $T$ are unbound.
\end{itemize}
\end{defin}

For $g=T(x_1,\ldots,x_n)$, $gr(g, A_1 \sqsubseteq A_2)$ is the atom $T'(x'_1,\ldots,x'_n)$, where
\begin{itemize}
\item $T'=\M(A_1)$, $T=\M(A_2)$;
\item the variables in $T'(x'_1,\ldots,x'_n)$ that correspond to the only component of $T'$ are equal to the ones that correspond to the only component of $T$;
\item the remaining variables in $T'(x'_1,\ldots,x'_n)$ are fresh.
\end{itemize}

\begin{defin}
Given an atom $g_1=r(X_1,\ldots,X_n)$ and an atom $g_2=r(Y_1,\ldots,Y_n)$, we say that $g_1$ and $g_2$ unify if there exists a variable substitution $\theta$ such that $\theta(g_1)=\theta(g_2)$. Each such a $\theta$ is called unifier. Moreover, if $g_1$ and $g_2$ unify, we denote as \texttt{mgu}$(g_1,g_2)$ a most general unifier of $g_1$ and $g_2$.\\
\end{defin}
We are now ready to define the algorithm \texttt{Rewrite}.

\paragraph{Algorithm}
At first, SQL query is translated to conjunctive query using standard SQL-to-CQ algorithm. Then the \texttt{Rewrite} algorithm is applied. Note, that the order of the variables, which is the one given by the translation from SQL to CQ, must be considered. 

\begin{verse}
\textbf{algorithm} \texttt{Rewrite}($q, \ \S$)\\
\textbf{input:} conjunctive query $q$, \dlrdb system $\system$ \\
\textbf{output:} union of conjunctive queries $P$\\
$P := \{q\};$\\
\textbf{repeat\\ 
\quad $P' := P;$\\
\quad for each $q \in P'$ do\\
\quad (a) for each $g \ in \ q$ do\\
\qquad for each $I \ in \ \K$ do\\
\qquad \qquad if $I \ is \ applicable \ to \ g$\\ 
\qquad \qquad then $P:=P \cup q[g/gr(g, I)]$\\
\quad (b) for each $g_1, \ g_2 \ in \ q$ do\\
\qquad if $g_1 \ and \ g_2$ unify\\
\qquad then $P:=P\cup\{\texttt{reduce}(q,g_1,g_2)\};$\\
until $P'=P$;\\
return $P$
}
\end{verse}

In the algorithm, $q[g/g']$ denotes the query obtained from $q$ by replacing the atom $g$ with a new atom $g'$.

Informally, the algorithm \texttt{Rewrite} first reformulates the atoms of each query $q \in P'$ and produces a new query for each atom reformulation (step (a))[5] . More precisely, if there exists an inclusion assertion $I$ and a conjunctive query $q \in P'$ containing an atom $g$, then the algorithm adds to $P'$ the query obtained from $q$ by replacing $g$ with $gr(g, I)$. 
For the step (b), the algorithm \texttt{Rewrite} for each pair of atoms $g_1,g_2$, that unify, computes the query $q'=\texttt{reduce}(q,g_1,g_2)$, obtained from $q$ by the following algorithm: 

\begin{verse}
\textbf{algorithm \texttt{reduce}($q,g_1,g_2$)\\
input:} conjunctive query $q$, atoms $g_1,g_2 \in body(q)$\\
\textbf{output:} reduced conjunctive query $q'$ \\
$\quad q':=q;$\\
$\quad \sigma := \textrm{mgu}(g_1,g_2)$\\
$\quad body(q'):=body(q')-\left\{g_2\right\}$\\
$\quad q':=\sigma(q')$\\
\textbf{return} $q'$\\
\end{verse}

Informally, the algorithm \texttt{reduce} starts by eliminating $g_2$ from the query body; then the substitution $mgu(g_1, g_2)$ is applied to the whole query (both the head and the body).

In order to compute the answers of $q$ to $\S$, we need to evaluate the set of conjunctive queries $P$ produced by the algorithm \texttt{Rewrite}. Every query $q$ in $P$ is transformed into an SQL query. The algorithm \texttt{Answer}, given a satisfiable KB $\K$ and a query $q$, computes the answer to $q$ over $\K$. $\texttt{Eval}(q,\D)$ denotes the evaluation of the SQL query $q$ over the database $\D$.

\begin{verse}
\textbf{algorithm} \texttt{Answer}($q, \S, \D$)\\
\textbf{input:} conceptual query $q$, \dlrdb system $\system$, database $\D$ for $\R$ \\
\textbf{output:} $ ans(q, \S, \D)$ \\
$\K$:=\texttt{Normalize}($\K$);\\
\textbf{return} \texttt{Eval}(SQL(\texttt{Rewrite}($CQ(q, \S),\S), \R), \D)$\\
\end{verse}

\section{Conclusions}

In this document we have described \dlrdb, a query answering system that enables to pose queries over the conceptual schema of a database, re-expressing a conceptual query in terms of relations at the underlying database and evaluating the rewriting over the underlying database. We have extended the DL \dllite to the DL \dlrlite which supports n-ary relations, without loosing nice computational properties of the developed reasoning techniques.

These results are advantageous in formulating more simple queries, using terms defined in the conceptual schema only, and infering additional information that was not stated explicitly in the user query but is implied by the constraints at the conceptual level. At the same time, the formalization of the conceptual schema and the reasoning techniques allow for checking the consistency of the underlying database against the conceptual schema.

\nocite{CaLR03b,LeLR04,BCMNP03,CaDL98}

\bibliographystyle{abbrv}
\bibliography{main}

\end{document}